\documentclass[apj,twocolumn]{emulateapj}
\slugcomment{Version: \today}

\begin{document}

\author{Gudlaugur J\'ohannesson, Gunnlaugur Bj\"ornsson and Einar H.
	Gudmundsson\altaffilmark{1}}
\title{Afterglow Light Curves and Broken Power Laws: A Statistical
	Study}

\altaffiltext{1}{Science Institute, University of Iceland, Dunhaga~3,
	IS--107 Reykjavik, Iceland, e-mail:gudlaugu@raunvis.hi.is,
	gulli@raunvis.hi.is, einar@raunvis.hi.is}

\begin{abstract}
  In gamma-ray burst research it is quite common to fit the afterglow
	light curves with a broken power law to interpret the data.  We apply
	this method to a computer simulated population of afterglows and find
	systematic differences between the known model parameters of the
	population and the ones derived from the power law fits.  In general,
	the slope of the electron energy distribution is overestimated from
	the pre-break light curve slope while being underestimated from the
	post-break slope.  We also find that the jet opening angle derived
	from the fits is overestimated in narrow jets and underestimated in
	wider ones.  Results from fitting afterglow light curves with broken
	power laws must therefore be interpreted with caution since the
	uncertainties in the derived parameters might be larger than estimated
	from the fit.  This may have implications for Hubble diagrams
	constructed using gamma-ray burst data.
\end{abstract}


\section{Introduction}
Many gamma-ray burst (GRB) afterglows can be interpreted within the
standard fireball model \citep[see][for a recent review]{Piran2005},
where a jet structure is implied from the steepening of the light curve
\citep{Rhoads1999}.  By fitting the light curves with a (sharp or
smoothly joined) broken power law \citep[see e.g.][]{Beuermann1999}, it
is possible to determine the time of the steepening (the so called jet
break time) and the pre-break and post-break slopes of the light curves.
This data can then be used to obtain information about the parameters of
the underlying model, e.g. the jet opening angle which can then be used
to estimate the energy of the burst
\citep[e.g.][]{Frail2001,Ghirlanda2004, Friedman2005}.

Although it has been shown that the afterglow light curves in the
standard model can be approximated with a broken power law
\citep[e.g.][]{SarPirNar1998,Rhoads1999}, the accuracy of the model
parameters derived in that way has not been tested.  In this letter we
show that the results from fitting a broken power law to afterglow light
curves must be interpreted with caution as the parameters derived from
the fit can be systematically different from the actual model
parameters.  We do this by creating a population of simulated
afterglow light curves using our version of the standard model
\citep{Johannesson2006} and then fit them with a broken power law.  In
section 2 we shortly describe the model and procedure used, present the
results and then conclude in section 3 with a discussion of our
findings.

\section{The model and results}

\begin{figure}
	\plotone{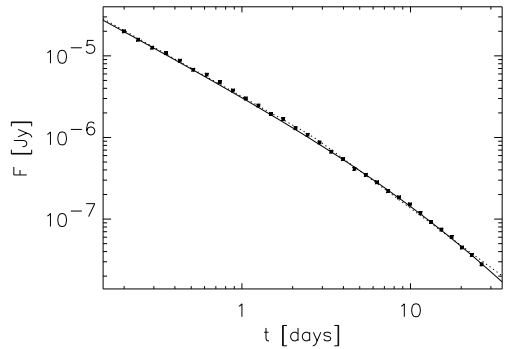}
	\caption{A typical light curve from our sample.  The points are from
	our model calculations, the solid curve shows the smooth fit and the
	dotted curve the sharp fit.  The light curve break time determined by
	the smooth fit is rather accurate in this case, around 22 days,
	whereas the sharp fit break time is around 3 days.  The light curve
	slopes agree well with the data for the smooth fit but the post-break
	slope in the sharp fit is too shallow as may be expected from the
	break time error.}
	\label{fig:exlc}
\end{figure}

We used the instantaneous energy injection version of the standard model 
presented in \citet{Johannesson2006}.  This 
is an extension of the \citet{Rhoads1999} model, and
includes a more detailed calculation of the synchrotron emission and the
effects of the equal arrival time surface (EATS).  We created a population of
20,000 afterglow light curves where the model parameter values were
selected at random from a normal distribution over a narrow parameter
range given in table~\ref{tab:parameterrange}.  We assume a constant
density interstellar medium for all events.  The standard deviation of
the distribution for each parameter was fixed at 1/4 of its range and
the distribution was clipped to fit within each parameter range.  The
range of parameters was chosen after fitting several afterglows with our
model over a wide range of frequencies \citep[see examples
in][]{Johannesson2006, Postigo2005}.  Although the fraction of energy
contained in electrons, $\epsilon_e$, and magnetic field,
$\epsilon_B$, do not directly enter our formalism, they are included as
they can affect the results via the characteristic synchrotron
frequencies, $\nu_m$ and $\nu_c$ \citep{SarPirNar1998}.  The viewing angle of the observer, $\theta_v$,
can smooth the jet break \citep{Granot2001} and was also
included as a parameter.  The light curves were evenly sampled in the
logarithm of observer time and consisted of about 35 points each (see
fig.~\ref{fig:exlc} for a typical example).  To make the synthetic
data more realistic, we added normally distributed random fluctuations
with a standard deviation of 3\%.  We also assumed a fixed error of 5\%
for each data point.  We find that our results are not sensitive to the
values of these error parameters and lowering the error estimate only
reduces the number of successful fits without reducing the scatter in
our results.  The results were also tested for the effects of the EATS 
by turning it off in a test sample.  We found that
it had no significant effect.

\begin{table}
	\centering
	\caption{The parameters of the afterglow population.}
	\label{tab:parameterrange}
	\begin{tabular}{lcc}
		Parameter           & Min Value & Max Value\\
		\tableline
		$E_0/10^{51}$ ergs  & 0.5       & 2\\
		$\Gamma_0$          & 100       & 1000\\
		$n_0$ cm$^{-3}$     & 0.1       & 10\\
		$\theta_0$          & 1\degr    & 20\degr\\
		$p$                 & 2         & 2.6\\
		$\epsilon_e$        & 0.1       & 0.5\\
		$\epsilon_B/10^{-4}$& 0.5       & 20\\
		$\theta_v/\theta_0$ & 0         & 1\\
		$z$                 & 0.1       & 6\\
	\end{tabular}

	{The parameters in the afterglow population are normally distributed
	over the ranges shown here.  The parameters are all distributed
	logarithmically, except $p$, $\theta_v/\theta_0$ and $z$.  See text
	for more detail.}
\end{table}

According to standard fireball theory, the simplest afterglow light
curves can be described with the widely adopted analytical approximation
\citep{Rhoads1999},
\begin{equation}
	\label{eq:brapp}
	F_\nu(t) \propto 
	\left\{
	\begin{array}{ccc}
		t^{-3(p-1)/4} & t< t_j,\\
		t^{-p} & t > t_j,
	\end{array}
	\right.
\end{equation}
for a given observing frequency, $\nu$, in the range $\nu_m < \nu <
\nu_c$.  Here $t$ is the time from the onset of the burst in the
observer's frame, $t_j$ is the jet break time in the observer's frame
and $p$ is the electron energy distribution index.  Since the jet break
is in general not sharp, eq.~(\ref{eq:brapp}) is often replaced
with a smoothly joined broken power law \citep{Beuermann1999},
\begin{equation}
	\label{eq:Beuermann}
	F_\nu(t) = F_j\left[\left(\frac{t}{t_j}\right)^{\alpha_1
	n}+\left(\frac{t}{t_j}\right)^{\alpha_2 n}\right]^{-1/n}.
\end{equation}
Here, $F_j$ is the flux at $t_j$, $-\alpha_1$ and $-\alpha_2$ are
respectively the pre-break and post-break light curve slopes and $n$ is
a numerical factor controlling the sharpness of the break.

The light curve break occurs when the jet enters the sideways expanding
regime \citep[called exponential regime in][]{Rhoads1999} and the
observer receives light from the entire jet surface.  The break time can
be approximated as that point in time when $\theta\approx 1/\Gamma$,
where $\theta$ is the opening angle of the jet and $\Gamma$ is the
Lorentz factor of the relativistically moving shock front.  Using
analytical approximations from \citet{Johannesson2006} we find that
\begin{equation}
	\label{eq:t_jet}
	t_j \approx 1.21(1+z)\left( \frac{E_0/10^{51}\,{\rm
	ergs}}{n_0/1\,{\rm cm}^{-3}} \right)^{1/3}
	\left(\frac{\theta_0}{0.1}\right)^2\quad{\rm days},
\end{equation}
where $E_0$ is the total energy injected into the jet, $n_0$ is the
constant interstellar medium particle density, $\theta_0$ is the initial
opening angle of the jet in radians and $z$ is the redshift of the
burst.  This formulation differs slightly from the one given in
\citet{Rhoads1999}, because we choose to use the energy injected into
the jet rather than the isotropic equivalent energy in order to better
isolate $\theta_0$ in the equation.

As is often done when fitting GRB afterglows
\citep[e.g.][]{Wijers1997,Beuermann1999,Zeh2005}, we fitted the
synthetic R-band light curves with both a sharply broken power law as in
equation~(\ref{eq:brapp}) and a smoothly joined broken power law as in
equation~(\ref{eq:Beuermann}).  These will hereafter be referred to as
sharp fits and smooth fits, respectively.  The Levenberg-Marquardt
method of \citet{nr} was used to minimize the $\chi^2$ value of the fit
in both cases.  Although the starting point of the fitting procedure was
chosen as the theoretically correct values of $t_j$, $\alpha_1$,
$\alpha_2$ and $F_j$, it did not result in an acceptable fit in every
case.  Only those events where the $\chi^2$ per degree of freedom is
less than 1 for the smooth fits and 2 for the sharp fits were selected
for further study.  A higher threshold was used for the sharp fits since
the numerically generated light curves are very smooth and not well
represented by a sharply broken power law (see figure~\ref{fig:exlc}).
About half of the fits fulfilled those requirements in both cases.
Since GRB afterglow measurements normally extend from a few hours after
the burst to about a month, we limited our data and $t_j$ to this time
range in the fitting procedure.  Using equation~(\ref{eq:t_jet}), it can
be shown that this range also limits the range of derivable opening
angles to about 1.5\degr - 15\degr, depending on other burst parameters.
With real data, this range of opening angles can be further reduced if
there is a bright underlying host or a supernova component making the
afterglow light difficult to observe.

\begin{table}
	\centering
	\caption{The parameters of the Gaussian fits to the error distributions.}
	\begin{tabular}{lcccc}
		 & \multicolumn{2}{c}{Smooth} & \multicolumn{2}{c}{Sharp} \\
		 & Mean & Std. dev. & Mean &  Std. dev. \\
		\tableline
		$\Delta p_1/p$ & 0.097 & 0.042 & 0.13 & 0.042 \\
		$\Delta p_2/p$ & -0.18 & 0.064 & -0.29 & 0.068 \\
		$\Delta \theta_j/\theta_0$ & -0.034 & 0.091\tablenotemark{*} & -0.19 &
		0.28\tablenotemark{**} \\
		\tablenotetext{*}{The Gaussian fit is not particularly good here and
		the distribution is actually wider.}
		\tablenotetext{**}{There is a narrow peak around 0 on top
		of the Gaussian.}
	\end{tabular}
	\label{tab:Gaussfit}
\end{table}

\begin{figure}
	\plotone{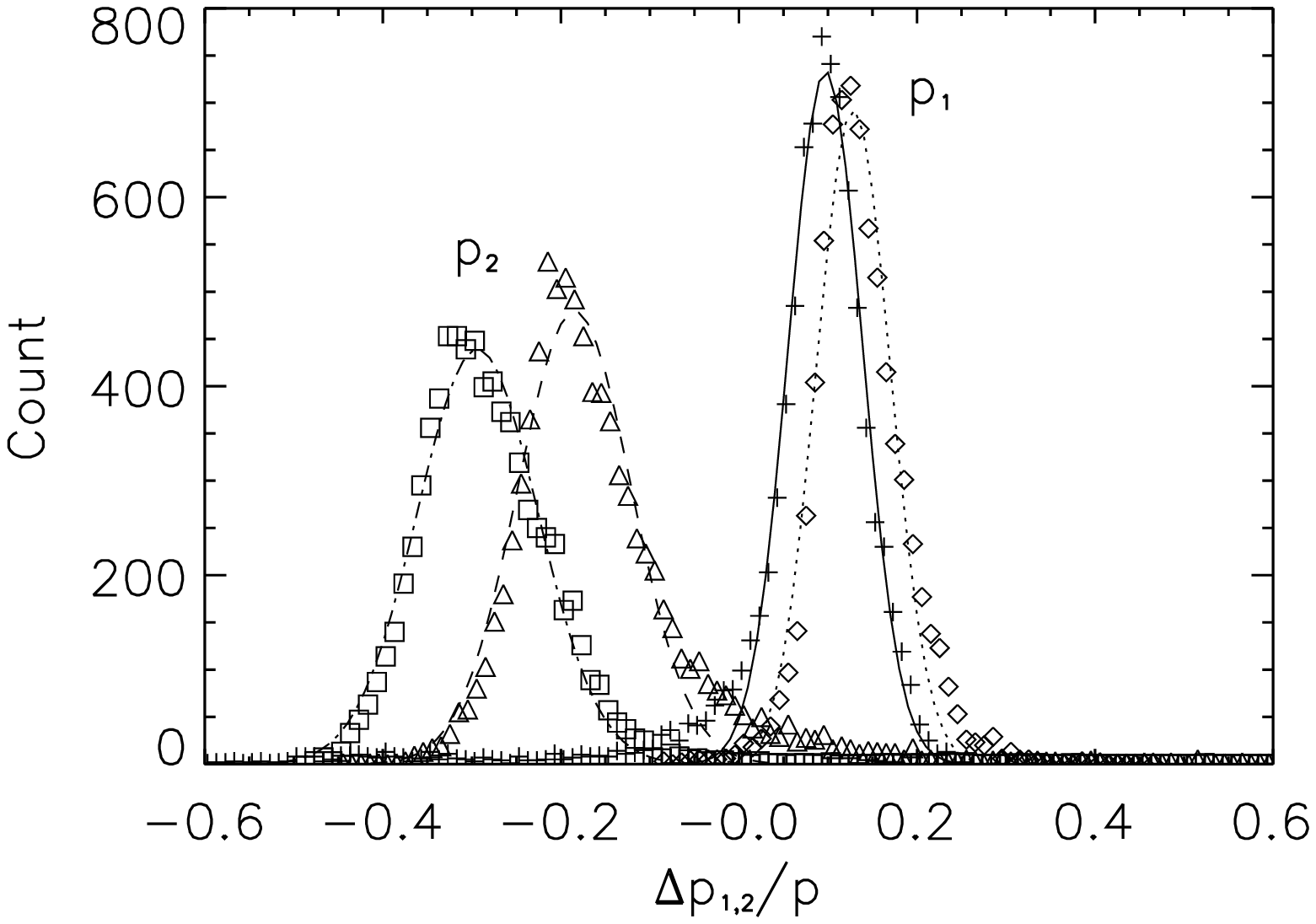}
	\plotone{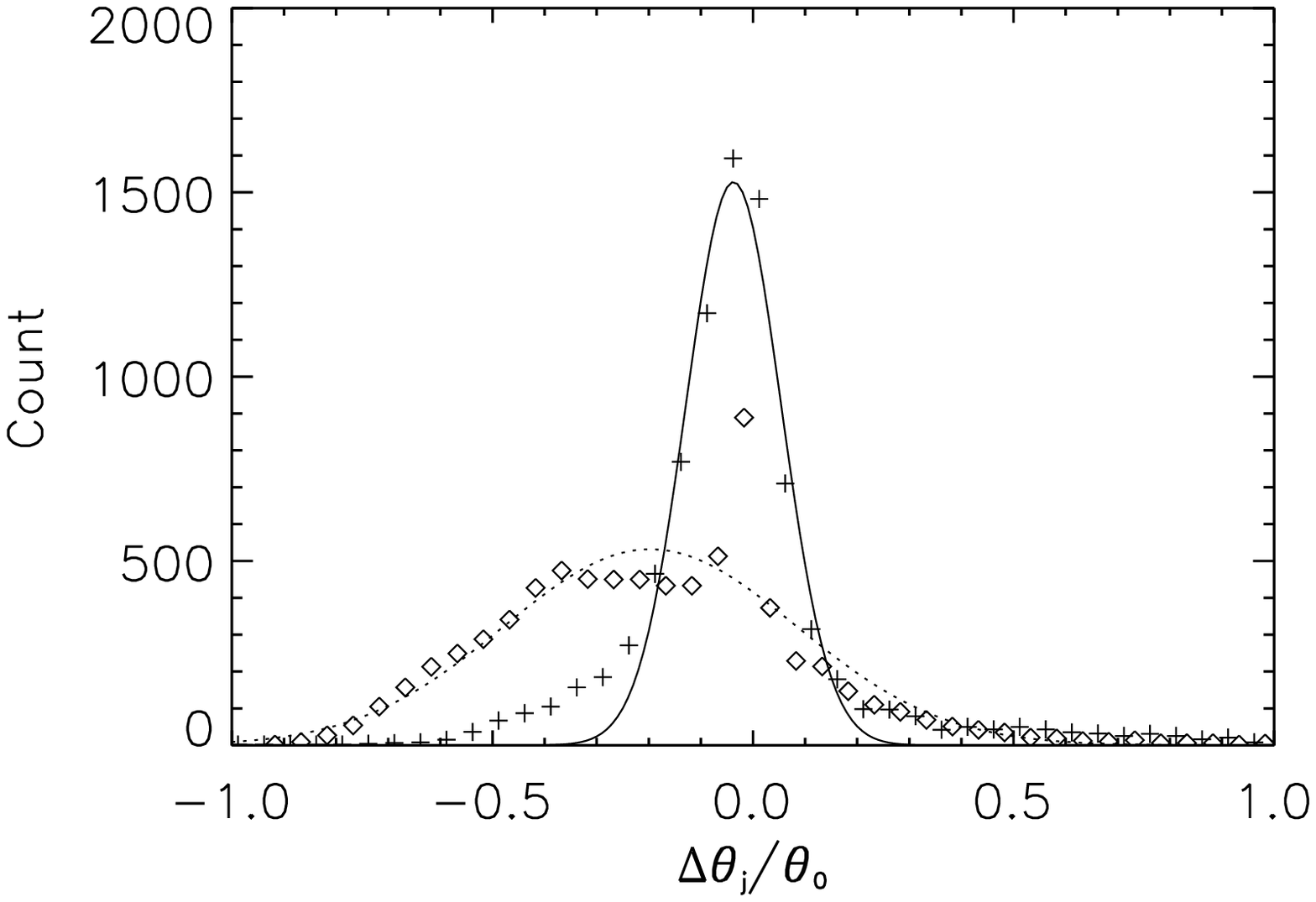}
	\caption{The distribution of relative differences in the value of
	parameters deduced from the fits.  The upper panel shows the
	distribution in $\Delta p_1/p$ and $\Delta p_2/p$.  The smooth and
	sharp fits are represented, respectively, by plus-signs and diamonds
	for $\Delta p_1/p$ and triangles and boxes for $\Delta p_2/p$.  The
	lower panel shows the distribution in $\Delta \theta_j/\theta_0$ for
	smooth (plus-signs) and sharp (diamonds) fits.  The overlaid curves
	are Gaussian fits to the corresponding distributions. The parameters
	of the Gaussians are given in table~\ref{tab:Gaussfit}.  The count is
	higher in the lower panel because the binsize is larger.}
	\label{fig:differences}
\end{figure}

The results for $t_j$ from the fitting procedure can be used to find an
estimate of the opening angle of the jet by inverting
equation~(\ref{eq:t_jet}),
\begin{equation}
	\label{eq:th_jet}
	\theta_j = \left(\frac{t_j/1\,{\rm day}}{121(1+z)} \right)^{1/2}\left(
	\frac{n_0/1\, {\rm
	cm}^{-3}}{E_0/10^{51}\,{\rm ergs}}
	\right)^{1/6}.
\end{equation}
Here, the opening angle is denoted by $\theta_j$ to distinguish it from
the known initial opening angle of the burst, $\theta_0$, from our
numerical model calculation, although theoretically they should be
equal.  The results for the derived slopes, $\alpha_1$ and $\alpha_2$,
can be used to find the value of $p$ and equation~(\ref{eq:brapp}) gives
two different values, $p_1 = (4/3)\alpha_1+1$ and $p_2 = \alpha_2$.
Comparison of the values obtained this way from the fitting procedure to
the parameters known from the numerical model is shown in
figure~\ref{fig:differences} as the distribution of relative differences
between the derived parameters and the known model parameters: $\Delta
p_1/p$, $\Delta p_2/p$ and $\Delta \theta_j/\theta_0$, where $\Delta p_1
= p_1 - p$, $\Delta p_2 = p_2 - p$ and $\Delta \theta_j = \theta_j -
\theta_0$.  Those distributions were each fitted with a Gaussian profile
that is overlaid on the distributions in the figure.  The parameters of
the Gaussian profiles are presented in table~\ref{tab:Gaussfit}.

The distributions and the corresponding Gaussian fits clearly show a
systematic difference in the evaluation of $p$, where $p_1$ is
overestimated and $p_2$ underestimated.  From the $\Delta
\theta_j/\theta_0$ distribution, it is clear that the sharply broken
power law does not do a good job in determining the opening angle and
there is both a significant underestimate and a large standard
deviation.  This is also reflected in the $\Delta p_2/p$ distribution
since an underestimate of the opening angle will make $p_2$ smaller.  It
should however be noted that there is a high narrow peak around zero in
the $\Delta \theta_j/\theta_0$ distribution for the sharp fits which is
not included in the Gaussian fit.  The overall standard deviations in
the Gaussian fits are not very high but should be considered when using
$\theta_j$ to correct the isotropic energy, since the correction factor
is approximately proportional to $\theta_j^2$.

\begin{figure}
	\plotone{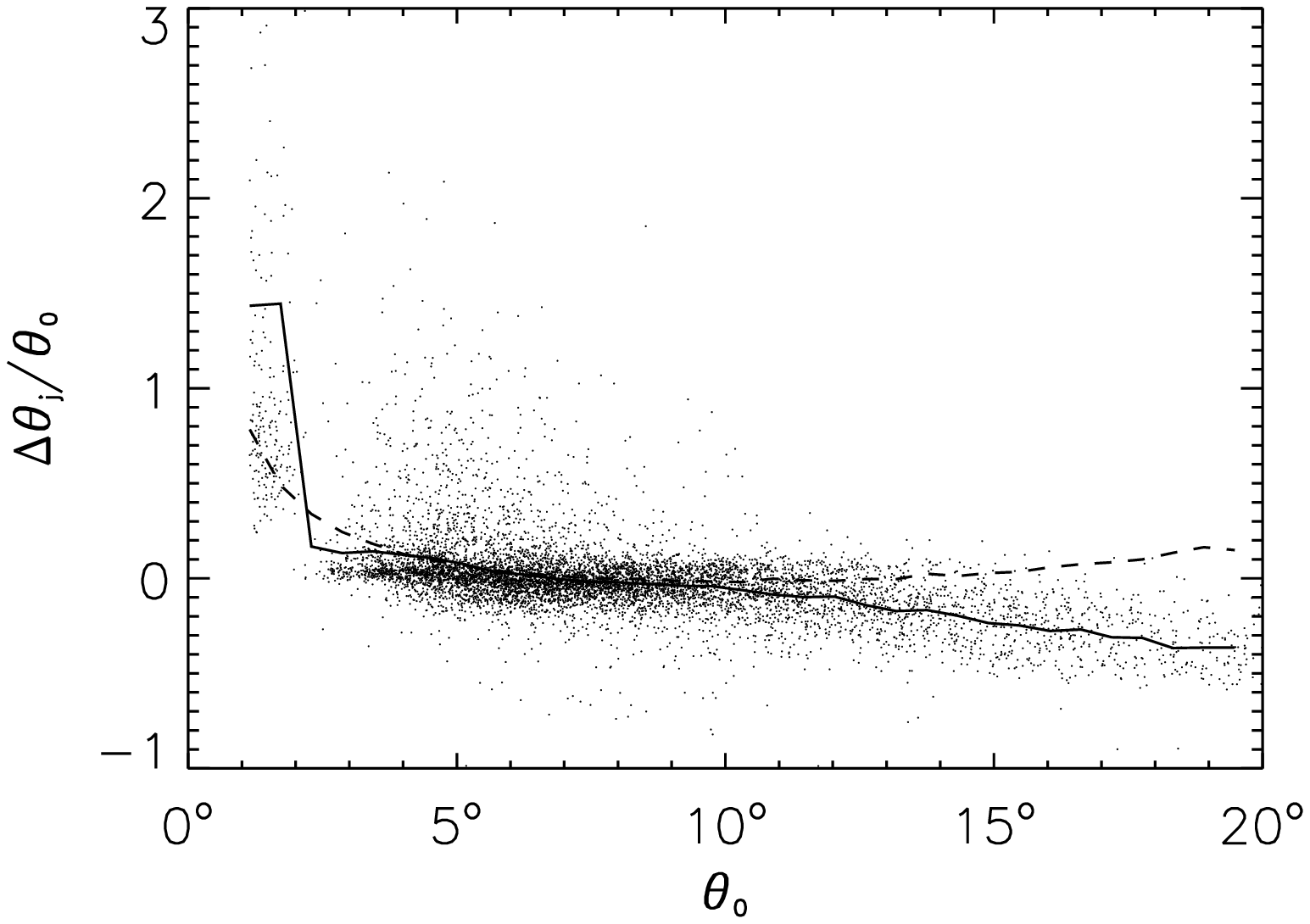}
	\plotone{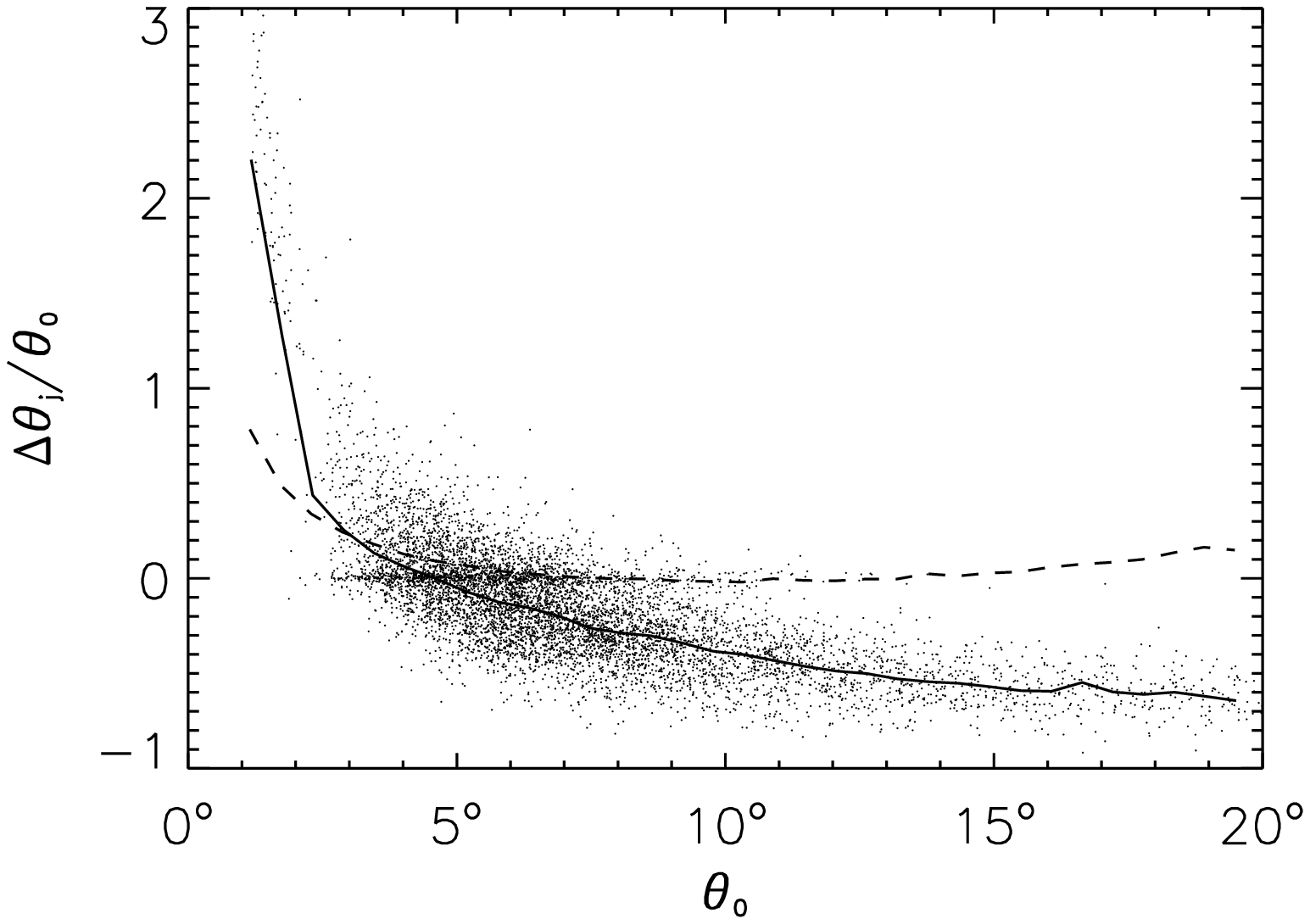}
	\caption{The correlation between $\theta_0$ and $\Delta
	\theta_j/\theta_0$ for smooth fits (top) and sharp fits (bottom).  In
	these scatter plots, the solid curves show a binned average of the
	data.   Also shown is the binned average calculated from
	$t_{j,\Gamma}$ (dashed curve).}
	\label{fig:dth0thj}
\end{figure}

We have checked for correlations between the parameters $p_1$, $p_2$ and
$\theta_j$ deduced from the fits and the known parameters of our
numerical model, which range is shown in table~\ref{tab:parameterrange}.
We find no significant correlation within that parameter range, except
between $\theta_0$ and $\Delta \theta_j/\theta_0$.
Figure~\ref{fig:dth0thj} shows a scatter plot of $\Delta
\theta_j/\theta_0$ as a function of $\theta_0$ for both smooth and sharp
fits and also a binned average shown as a solid line.  It is clear that
narrow jet opening angles are systematically overestimated and wider
ones underestimated.  This is more evident with the sharp than the
smooth fits.  When the jet opening angle is used to correct the
isotropic energy for beaming, this will lead to a clustering of the
derived energy values.

To check the validity of the approximations used in deriving
equations~(\ref{eq:t_jet})~and~(\ref{eq:th_jet}) we determined the
characteristic time $t_{j,\Gamma}$ when $\theta$ equals $1/\Gamma$ in
our model calculations.  We then estimated the opening angle
by setting $t_j$ equal to $t_{j,\Gamma}$ in equation~(\ref{eq:th_jet}).
This should give correct results in each case if these approximations
are valid.  The resulting angle, $\theta_j$, is then compared to the
known model opening angle, $\theta_0$, in a similar way as the for the
power law fits.  Figure~\ref{fig:dth0thj}  also shows the binned average
of the correlation between $\Delta \theta_j/\theta_0$ and $\theta_0$ for
$\theta_j$ calculated from $t_{j,\Gamma}$ (dashed curve).  It clearly
shows that the approximations used in deriving
equations~(\ref{eq:t_jet})~and~(\ref{eq:th_jet}) break down for very
narrow jets.  This is because the jet break time can be very close to
the deceleration time \citep{Panaitescu1998} where the approximations
are not valid.  These approximations also break down in very wide
jets since these may not necessarily be assumed to be ultra relativistic
at the time of the jet break.  Figure~\ref{fig:dth0thj} also shows that
the correlation is stronger when $\theta_j$ is calculated from $t_j$
(solid curves), so $t_j$ is overestimated from the fits when the jet is
narrow and underestimated in wide ones.  One explanation is that the jet
break time is not necessarily within the limited time range used in the
fits, but it could also be due to $\nu_m$ crossing the observing
frequency, especially in narrow jets.  In the latter case, those events
can be identified from a small value of the smoothness factor, $n$,
because the break around $\nu_m$ is much smoother than the jet break.
This crossing of $\nu_m$ through the observing frequency is also the
cause of the small wings in the distribution of $\Delta p_1/p$ and
$\Delta p_2/p$ seen around the value -0.1 in
figure~\ref{fig:differences}.  By looking at the individual spectra, 
we can identify those events where $\nu_m$ or
$\nu_c$ cross the observing frequency.  We find that the latter case
does not have a significant effect on our results.

\section{Conclusion}

We have shown that results from the standard procedure of fitting
afterglow light curves with a broken power law must be interpreted with
caution.  There can be systematic differences in the evaluation of the
electron energy distribution index from the slope of the fitted curves
and a strong correlation between the relative difference in the opening
angle estimated from the light curve break time and the initial opening
angle of the jet.  These findings are partly due to the approximations
used in deriving equations (1)-(4) being used out of their validity
limits.  This applies to approximations in both the dynamical
\citep{Bianco2005} and the radiation properties of the expanding shell.
These differences and particularly the correlation is also a consequence
of difficulties in accurately determining the jet break time from the
afterglow light curves.

The fitting procedure we used was completely automatic and often did not
converge.  In some cases it resulted in erroneous parameter values and
we therefore adopted a threshold on the reduced $\chi^2$ to eliminate
those bad fits.  To test the effect of this threshold, we removed it
from the selection criteria and re-examined the data, still considering
only those events where the light curve break time was within the time
range of our data.  This left us with over 90\% of the original
population.  The most significant changes we find in the results, were
slightly larger standard deviations in the relative difference
distributions and more spread in the $\Delta \theta_j/\theta_0$ -
$\theta_0$ correlation thereby weakening it.  The spike around 0 seen in
the $\Delta \theta_j/\theta_0$ distribution for the sharp fits in
figure~\ref{fig:differences} (diamonds) also becomes a dominating
feature.  This indicates that by using the $\chi^2$ as the strongest
filter, we remove the bad fits from the ensemble but may also lose some
useful fits.

It is known theoretically that the magnitude of the jet break, $\Delta
\alpha = \alpha_2 - \alpha_1$, can be used to differentiate between a
wind like or a constant density environment
\citep[e.g.][]{Panaitescu1998}.  The systematic differences we find in
the evaluation of $p$ indicate that $\Delta \alpha$ is underestimated in
most of our events.  Similar conclusion is obtained for a population of
afterglow light curves in a wind medium, the main difference being that
$p_1$ did not show a systematic deviation in that case.  This renders
the method of using $\Delta \alpha$ to distinguish between density
profiles impractical.

The correlation between the opening angle estimate and the known opening
angle puts a strong limit on interpretations of the beaming corrected
energy of the burst \citep[e.g.][]{Frail2001, Ghirlanda2004,
Friedman2005}.  The clustering of the jet break time due to limited time
span of the data together with a generous use of the approximations used
in determining $\theta_j$ results in a bias towards moderate opening
angles, approximately between 2\degr - 10\degr.  The rather large
standard deviation in the $\Delta \theta_j/\theta_0$ distribution also
makes the results unreliable.  Using the beaming corrected energy as a
basis for cosmological studies therefore calls for a very careful
determination of $t_j$ and $\theta_j$.

The synthetic burst population studied in this paper is computer
generated and the power law fit should in theory be perfect.  The fact
that the differences between the parameters derived from the fits and
the known model parameters are so significant, makes the accuracy of
power law fits to real measurements of afterglow light curves a concern.
In real bursts, effects such as density fluctuations and energy
injection, can change the shape of afterglow light curves as may for
example be the case for GRB 021004 \citep[e.g.][]{Lazzati2002} and GRB
030329 \citep[e.g.][]{Sheth2003}.  These were densely sampled and were
not well fitted with a broken power law due to bumps in the light
curves.  It is not hard to see that grainier measurements of the same
events could have been fitted with a broken power law, leading to even
larger uncertainties in the parameter estimates than discussed here.
It should also be noted that other models are capable of explaining
light curve breaks, the most popular being the structured jet model
\citep{Rossi2002}.  There the light curve break depends on the
observer's viewing angle rather than the jet opening angle.  Hence the
methodology adopted in this letter is not directly applicable to that
model.

\acknowledgements
{We thank P.\ Jakobsson for critical comments on the manuscript.
This work was supported in part by a Special Grant from the Icelandic
Research Council, by the University of Iceland Research Fund and by the
Graduate Research Fund of The Icelandic Centre for Research.}

\end{document}